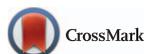

PAPER

OPEN ACCESS

# Light absorption enhancement of perovskite solar cells by a modified anti-reflection layer with corrugated void-like nanostructure using finite difference time domain methods

Budi Mulyanti[1,2,*] ⓘ, Muhammad Raihan Anwar[1], Chandra Wulandari[3,4], Lilik Hasanah[3], Roer Eka Pawinanto[1] ⓘ, Ida Hamidah[2,5] and Andrivo Rusydi[6,7,8]

[1] Department of Electrical Engineering Education, Universitas Pendidikan Indonesia, Jl. Dr Setiabudhi 229, Bandung 40154, Jawa Barat, Indonesia
[2] TVET Research Center, Universitas Pendidikan Indonesia, Jl. Dr Setiabudhi 229 Bandung 40154, Jawa Barat, Indonesia
[3] Department Physics Education, Universitas Pendidikan Indonesia, Jl. Dr Setiabudhi 229 Bandung 40154, Jawa Barat, Indonesia
[4] Engineering Physics, Faculty of Technological Industry, Institut Teknologi Bandung, Jl. Ganesha 10, Bandung, 40132, Jawa Barat, Indonesia
[5] Department of Mechanical Engineering Education, Universitas Pendidikan Indonesia, Jl. Dr Setiabudhi 229 Bandung 40154, Jawa Barat, Indonesia
[6] NanoCore, National University of Singapore, Singapore 117576, Singapore
[7] Department of Physics, National University of Singapore, 2 Science Drive 3, Singapore 117542, Singapore
[8] Singapore Synchrotron Light Source, National University of Singapore, 5 Research Link, Singapore 117603, Singapore
[*] Author to whom any correspondence should be addressed.

E-mail: bmulyanti@upi.edu

## Abstract

Perovskite solar cells (PSC) have become a growing research interest due to their flexibility, attractive properties, and low production cost. However, the thin-film structure of PSC often results in a not fully absorbed incident light by the active layer, which is crucial to determine PSC efficiency. Thus, the fabrication of an active layer with unique nanostructures is often used to enhance light absorption and general PSC efficiency. Using the theoretical simulation based-on Finite-Difference Time-Domain (FDTD) technique, this work demonstrates the successful improvement of light absorption by embedding corrugated void-like structure and perovskite thickness modification. The investigation of a corrugated void-type anti-reflection layer effect on light absorption is done by modifying the radius ($r$) and lattice constant ($a$) to obtain the optimum geometry. In addition, the MAPbI$_3$ perovskite layer thickness is also adjusted to examine the optimum light absorption within the visible length to near-infrared. The theoretical calculations show that the optimum $r = 692$ nm and $a = 776$ nm. Meanwhile, the optimum absorber layer thickness is 750 nm. Compared to flat PSC, our proposed PSC absorbed more light, especially in the near-infrared region. Our result shows demonstrates the successful enhancement of light absorption by embedding corrugated void-like structure and modifying the perovskite thickness using a theoretical simulation based on the FDTD technique.

## 1. Introduction

The increasing demand for renewable energy sources has occurred in the past decades due to their capabilities to provide reliable power supplies and fuel diversification [1]. Besides, adopting renewable energy sources helps to lessen the greenhouse effect, which is critical to alleviating environmental concerns [2, 3]. Photovoltaics stand out among the many renewable energy sources examined because of their many benefits, including being a clean energy source that can be used on an industrial scale or even in a single home [4, 5]. Based on production costs and physical qualities, photovoltaic systems can be classified as either first, second, or third generation [6]. Perovskite solar cells, also known as PSCs, are the third generation of photovoltaics and are considered an emerging technology. Some of the many advantages of this technology include its flexibility, tailored form factors, light-weight, semi-transparency, and responsiveness to a broad spectrum of light wavelengths [7]. The power conversion efficiency (PCE) of PSC has increased in the past decade from 3.8% (in 2009) to 25.7% (in 2021) [8, 9]. Furthermore, the increase in PSC efficiency is also accompanied by a more affordable price and relatively easy fabrication than conventional silicon-based photovoltaic technology [10]. As a result, there is a growing research interest in PSC technology as reported in the previous report.

Numerous studies have explored various strategies in PSC development to achieve significant PCE enhancement. In general, PSC modification is carried out at two levels: material modification and wafer structure modification. Several studies have applied the defect passivation strategy on their perovskite and ETL interface to modulate interfacial energy band alignment and facilitate the crystallization of perovskites. The anion modifiers, such as guanidinium thiocyanate and guanidine acetate, are applied in the perovskite and $SnO_2$ interface [11]. Meanwhile for perovskite with PEDOT:PSS ETL, the poly(triarylamine) was sandwiched on their interfaces [12]. The two defect passivation strategies resulted in a PCE of 19.04% and 23.74%, respectively [11, 12]. In addition, alkali ion diffusions such as $Li^+$ and $K^+$ can also increase PCE by modulating the electric properties of perovskites [13, 14]. The $Li^+$ was incorporated into the perovskite and $SnO_2$ ETL interface by directly adding the LiOH into the $SnO_2$ colloidal dispersion solution. The $Li^+$ doping produced the carrier transport and extraction increment, resulting in a higher PCE of 21.31% compared to the control devices [13]. The $K^+$ doping can also be applied to the perovskite and $SnO_2$ interfaces, producing a high PCE of 21.18% [14].

The higher efficiency was achieved by several types of modified wafer structure of PSC, including single junction PSC at 25.7% [9], perovskite/perovskite tandem solar cell PCE at 26.4% [15], and perovskite/silicon tandem solar-cell PCEs at 29.15% [16]. The single junction PSC can achieve the highest efficiency by modifying the electron transport layer (ETL). With polyacrylic acid–stabilized $SnO_2$ quantum dots (paa-QD-$SnO_2$) on the compact $TiO_2$ as ETL, the light captured was enhanced and the nonradiative recombination at the ETL-perovskite interface was suppressed [9]. Another study found modification of $TiO_2$-based ETL using $SnO_2$ thin shell-protected Ag nanowires (Ag/$SnO_2$ NWs) [15] and ZnO nanocrystals [16]. The perovskite/perovskite tandem solar cell combines wide and narrow bandgap perovskite in one PSC structure. The ammonium-cation-passivated Pb–Sn perovskites overcame the short carrier diffusion length problem in the Pb–Sn narrow-band gap subcell. Therefore, enabling the application of thick subcells (∼1.2 $\mu$m) to achieve high photocurrent density in tandem solar cells [17]. The application of inverted p-i-n in tandem PSC also reported to provide efficiency enhancement for ∼3% compared to single junction PSC [18]. The recent study reported modifications in the form of dual ETL applications [19], front contact design [20], and perovskite band-engineering [21] as expected steps to increase the PCE of perovskite/perovskite tandem solar cells. Meanwhile, the highest efficiency of the perovskite/silicon tandem solar cell is reached by increasing the hole extraction obtained from a self-assembled, methyl-substituted carbazole monolayer as the hole-selective layer [22]. More effort in the field of PSC modification still needs to be made to obtain a high PCE and reliable structure.

The high efficiency of PSC is mainly attributed to excellent absorbance properties, longer charge carrier diffusion lengths, and stable charge transport properties [7]. The PSC also reacts to various light wavelengths, converting more Sunlight to electricity [23]. Therefore, light absorption is one of the most important factors to consider during the PSC manufacturing process. An anti-reflection layer is frequently employed to improve PSC light absorption by reducing reflectional loss and enhancing light presence on the active parts [24]. A pevious study in optical characteristics of an anti-reflection protection layer for PSC's fabricated in ambient air revealed that the use of an anti-reflection layer enhances the optical characteristics and efficiency of PSCs by reducing the haze effect and enhancing transmittance [25]. Meanwhile, a separate study utilizing silica nanosphere antireflection coatings indicated that the efficiency of PSCs increased from 14.81% to 15.82%, which significantly reduced broadband and wide-angle reflectance [26]. Recently, a group of researchers utilized an anti-reflection layer built on plasma-polymerized-fluorocarbon (PPFC) thin-film, which considerably improved the transmittance and stability of PSC's under high humidity, heat, and chemicals. The presence of PPFC-based anti-reflection hence increases the PSC's efficiency from 18.6% to 20.4% [27].

In addition to the anti-reflection layer, PSC's efficiency was increased by optimizing the perovskite layer as an active part that absorbs light. According to a study on the light-intensity and thickness-dependent efficiency of planar PSC's, various PSC thicknesses result in distinct light absorption characteristics [28]. Higher efficiency is produced by a thicker perovskite layer at low light intensities (0.1 to 0.5 Sun), making it appropriate for functioning in low light conditions comparable to those found in practical applications. Meanwhile, at high light intensity (> 1 Sun), a thinner perovskite layer produces higher efficiency. Increasing the thickness of the perovskite layer improves light absorption, decreases interfacial recombination, and increases biomolecular recombination losses. Reduced perovskite layer thickness, on the other hand, exhibits the contrast trend and is suitable for PSC under intense illumination. Therefore, it is essential to optimize PSC's thickness in order to achieve the optimum results.

The present work demonstrates the theoretical design simulation of PSC by combining corrugated anti-reflection layer and perovskite layer thickness optimization. Simulation work is required to predict physical behavior and the output performance of PSC. This work will be beneficial to predict the best parameters possible for the development of PSC. Firstly, the anti-reflection layer was designed to have a corrugated structure of a corrugated void-like structure, which is believed to help reduce light reflection and maximize the light distribution throughout the active layer [23, 29, 30]. The values for radius (*r*) and lattice constant (*a*) of the corrugated void-like structure of the anti-reflection layer were varied to optimize its performance. Then, the thickness of the perovskite layer was also optimized to find what thickness performs the best light absorption behavior. Afterward, the finite-difference time-domain (FDTD) method was used to investigate the behavior of the light propagation throughout the entire PSC model. Therefore, several related phenomena, including light absorption at the anti-reflection and perovskite layers, can be calculated to observe the best combination of parameters to design the PSC. These results highlight the promising design potential for developing PSC with corrugated void-like structure and an optimized perovskite layer thickness.

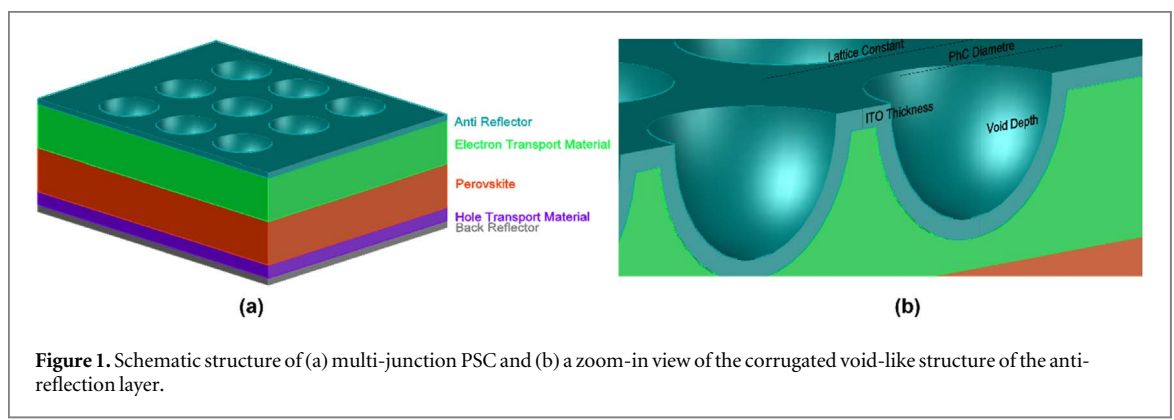

**Figure 1.** Schematic structure of (a) multi-junction PSC and (b) a zoom-in view of the corrugated void-like structure of the anti-reflection layer.

We construct the anti-reflection layer with a corrugated structure of a corrugated void-like structure to minimize light reflection and maximize light distribution throughout the active layer [23, 29, 30]. The anti-reflection layer's corrugated void-like structure radius (*r*) and lattice constant (*a*) values are varied to optimize its performance. The thickness of the perovskite layer is then altered to determine the optimal thickness for light absorption. In addition, the Finite-Difference Time-Domain (FDTD) method is employed to investigate the behavior of light propagation throughout the entire PSC model. Consequently, numerous related phenomena, including light absorption at the anti-reflection and perovskite layers, could be estimated to determine the optimal parameter combination for the PSC. These results indicate the viability of PSC with corrugated void-like structures and variable perovskite layer thickness.

## 2. Methods

Figure 1(a) shows the multi-junction PSC design structure studied in this work. The stacking order from the bottom to the top consists of a back reflector, a hole transport layer (HTL), a perovskite layer, an electron transport layer (ETL), and an anti-reflection layer. First, the back reflector layer material is silver (Ag) with a thickness of 80 nm. Next, the HTL material is Spiro-OMeTAD with a thickness of 150 nm. Then, the perovskite layer material is $CH_3NH_3PbI_3$ (MAPbI3) with a thickness of 500 nm for anti-reflection layer optimization and varied within 250 to 700 nm for perovskite thickness optimization. After that, the ETL material is titanium dioxide ($TiO_2$) with a thickness of 497.8 nm. Finally, the anti-reflection material is indium tin oxide (ITO) with a thickness of 62.3 nm. The material chosen as HTL and ETL refers to a study from Azri *et al* (2019) which compares the different transport layers [31]. Figure 1(b) shows the corrugated void-like structure studied in this work with the structure of a half-spherical-like void structure. For thickness layer optimization of the perovskite layer, the void parameters of depth and diameter and lattice constant are 467 and 388 nm, respectively. Meanwhile, the void's radius (*r*) and lattice constant (*a*) for the anti-reflection layer optimization varies within

An optical model is used to investigate the absorption in PSC devices. FDTD simulation provided by Lumerical Ltd used Maxwell equations to accurately solve the Maxwell equation related to the interaction between electromagnetic waves (light) and solar cells structure [32, 33]. The incident light bandwidth used in this study is 300–1000 nm based on the AM1.5 spectrum. Meanwhile, the photon flux at this bandwidth is high and low outside this range [33, 34]. In a solar cell, the power absorbed per unit volume in each material element is determined by the electric field resulting by the charge distribution from light illumination. According to the Poisson equations (equation (1)), this electric field can be described by the perturbation of charge transport and recombination [35].

$$\frac{\partial^2 \psi}{\partial^2 x} = -\frac{\partial E}{\partial x} = -\frac{\rho}{\varepsilon_s} = -\frac{q}{\varepsilon_s}[p - n + N_D^+(x) - N_A^-(x) \pm N_{def}(x)] \quad (1)$$

where, $\psi$ is the electrostatic potential, $q$ is elementary charge, $\varepsilon_s$ is the static relative permittivity of the medium, $n$ is the electron density, $p$ is the hole density, $N_D^+(x)$ is the density of ionized donors and $N_A^-(x)$ is the density of ionized acceptors, and $N_{def}$ is the possible defect acceptor and donor density [36].

Therefore, power absorbed per unit volume ($P_{Abs}$) is given by the equation (2), with $\omega$ is the angular frequency of light corresponding to the wavelength, $\varepsilon'$ is the imaginary part of the dielectric permittivity, and $|\vec{E}|^2$ is the electric field strength. Meanwhile, the light absorption for each wavelength is calculated by equation (3). $P_{Abs}(\lambda)$ is the power absorbed per unit volume for a given wavelength and dV corresponds to the volume of the absorber.

$$P_{Abs} = \frac{1}{2}\omega\varepsilon''|\vec{E}|^2 \quad (2)$$

$$Abs(\lambda) = \int P_{Abs}(\lambda)dV \quad (3)$$

FDTD simulation was performed on a 3-dimensional mesh cell of the PSC structure, with an accuracy of mesh 2. The light source uses a plane wave mode on the *y*-axis with amplitude and wavelength of 1 and 300–1000 nm, respectively. The *y*-axis uses a perfectly matched layer (PML) boundary condition (BC) to maximize the trapping of incident light. Meanwhile, periodic BC is applied to the *x*-axis with the assumption that the periodicity of the structure is infinite. A 2D y-normal frequency domain plane and a power monitor are placed to sandwich the specific layer to record absorption. The monitor is placed between the $TiO_2$ Absorber and the Spiro-OMeTAD Absorber for a general inspection of the structure. The 3D frequency-domain plane and power monitor were also used to record the electric field profile $|\vec{E}|^2$ with units of $V^2 m^{-2} Hz^{-1}$. Finally, this simulation can generate absorption curves and E-field profiles for the desired structures and layers.

## 3. Result and discussion

The PSC in this work has five different layers of ITO anti-reflection layer, $TiO_2$ as ETL, $MAPbI_3$ perovskite layer, Spiro-OMeTAD as HTL, and Ag back reflector, as shown in figure 1. The ITO anti-reflection layer is a conductive transparent substrate that can be used as a medium for the fabrication of PSC and the front contact of this device. Thus, the ITO anti-reflection layer should have good conductivity, high transmission, and good adhesion [37]. The $TiO_2$ ETL is a medium where the $MAPbI_3$ perovskite layer is deposited, creating a mesoporous interface for electron injection. The $TiO_2$ ETL has been widely used due to its low cost, suitable energy level, and high thermal stability [38]. Furthermore, the mesoporous structure provides a high surface area for perovskite material. The $MAPbI_3$ perovskite layer is PSC active parts that absorb light and generate electron-hole pairs. The electron-hole pairs were separated by the inner electric field of the $MAPbI_3$ perovskite layer. The electron moves to the $TiO_2$ ETL, and the hole moves to the Spiro-OMeTAD HTL. The Spiro-OMeTAD HTL covers the $MAPbI_3$ perovskite layer and collects the hole from the perovskite layer. Lastly, Ag is used as back-contact material for PSC.

At the beginning of the study, the anti-reflection layer was modified with a corrugated void-like structure. This modification was applied to introduce the light trapping phenomenon of the anti-reflection layer, which is believed to improve the absorption of the PSC [23, 29, 30]. Figure 2 shows the light absorption of the PSC with different void parameters. The variation of the void structure was done to understand how the structurization affects the absorption phenomenon. In this work, the parameters of the corrugated void-like structure were varied based on different lattice constant (*a*) and radius (*r*) of void structure, the influence of perovskite layer thickness is also studied.

The effect of the lattice constant on the PSC light absorption performance is hard to observe by looking at the entire photon energy range, therefore, the observations are made in a narrower range. Light absorption is high in the high energy range (2.0–4.0 eV), it tends to be constant at around 0.9 and has relatively the same absorption performance for each lattice constant variation. In the low energy range of 1.5–2.0 eV, the light absorption increases as the lattice constant decreases, namely 3880, 2587, and 1940 nm; in other words, absorption increases as the distance between the void structures increases. However, the difference was not seen significantly for a smaller size variation (776–1552 nm) because the size difference with each other was not significant. In lowest photon energy region <1.5 eV, the high absorption is dominated by PSC with the small void lattice constant, which is 776 nm. This absorption performance is supported by the light reflectance curve, which is very low in the high energy photon range. Meanwhile, at low energy photons, as observed in light absorption, the reflectance is also very fluctuating and is dominated by voids with *a* = 970 nm.

In the void radius variation, high absorption in the <1.5 eV region was produced by PSC with a void radius of 692 nm, which was very noticeable compared to other radius variations. Whereas in the broader area of 2.0–4.0 eV, a significant increase in absorption is provided by two small radius variations, namely 256 and 346 nm. These results generally agree with observations of variations in the lattice constant, where the greater the distance between the void structures, the higher the light absorption.

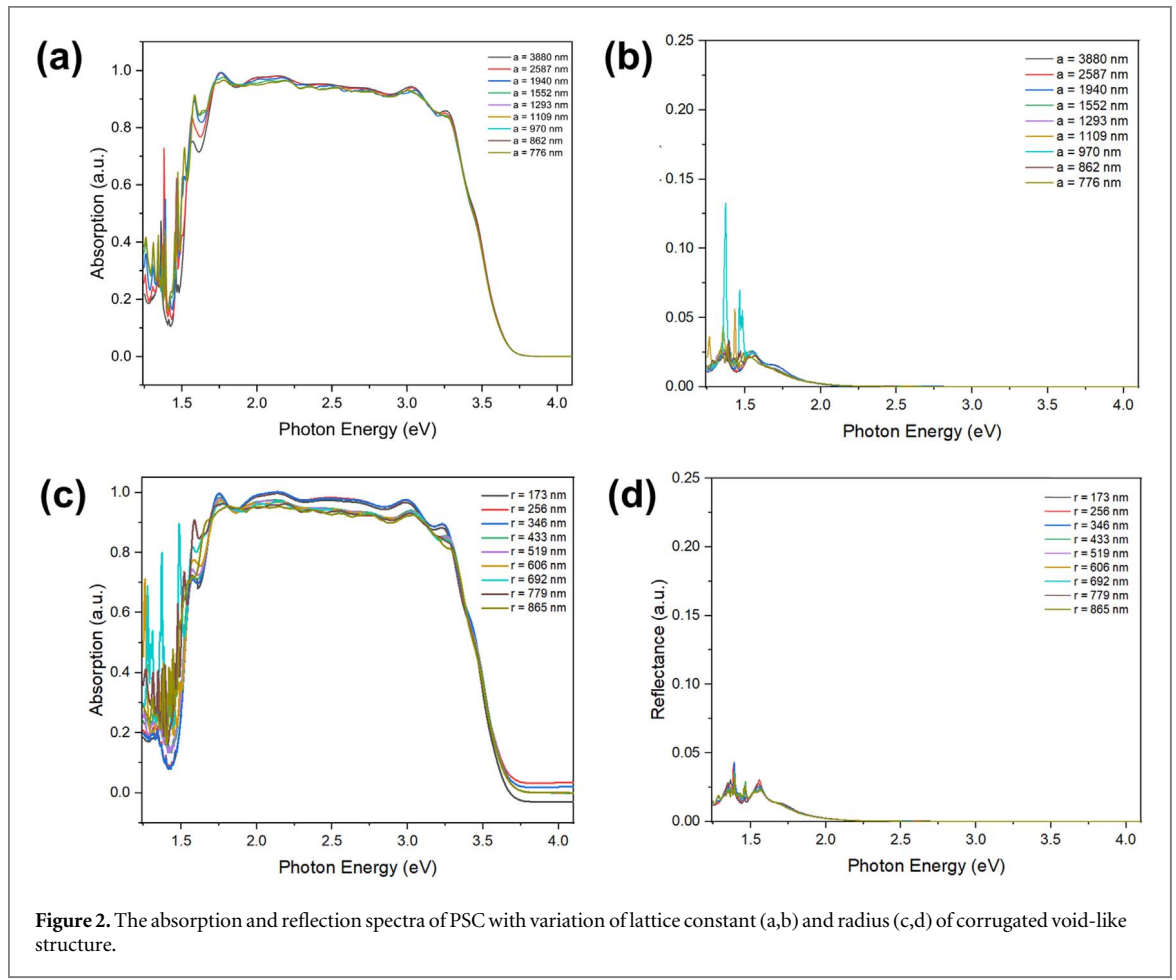

**Figure 2.** The absorption and reflection spectra of PSC with variation of lattice constant (a,b) and radius (c,d) of corrugated void-like structure.

The low performance of the light-trapping phenomenon on the anti-reflector layer occurs due to the lack of area to receive incoming photons. The increase in absorption is related to the absorption of light by the absorber assisted by the light trapping feature accompanied by minimal parasitic light reflection [39, 40]. This collaboration can provide a significant increase in the efficiency factor of PSC. In addition, this structuration also increases absorption at NIR wavelengths of solar wavelengths from 750 to around 1000 nm, which previously experienced a significant decrease in conventional PSCs [41].

The thickness of $MAPbI_3$ perovskite layer also plays an important role in optimizing PSC absorption since a trade-off exists between light absorption and carrier extraction [42]. A thick layer will increase series resistance and recombination of the $MAPbI_3$ perovskite layer, deteriorating the current collection efficiency. Meanwhile, a too-thin layer will reduce light absorption, damaging overall photocurrent production. Figure 3 shows the absorption spectra of $MAPbI_3$ perovskite layer with different thicknesses. For photon energy higher than 2.5 eV, the absorption spectra of every $MAPbI_3$ perovskite layer with different thicknesses seem to be sticking together, indicating that the thickness of the absorber layer does not significantly affect the absorption performance at those regions. At a below 2.5 eV, the absorption spectra of each $MAPbI_3$ perovskite layer with different thicknesses began to separate from each other. The average light absorption at photon energy lower than 2.5 eV increased as the thickness of $MAPbI_3$ perovskite layer increased. The thickness of 750 nm shows the best absorption compared to other thicknesses, indicating a large light capture area to maximize absorption. These results are consistent with the reflectance spectra which show high reflectance in PSCs with thin $MAPbI_3$ and low reflectance in thick $MAPbI_3$.

$MAPbI_3$ is a hybrid organic-inorganic material for the absorber layer in PSC, generally prepared via solutions process at 150 °C. $MAPbI_3$ has drawn a lot of attention due to its exceptional semiconducting properties and cost-effectiveness [42–45]. $MAPbI_3$ possesses a suitable bandgap of about 1.5 eV for absorber material in PSC application with a broad range absorption spectrum from the UV region to 825 nm. Furthermore, $MAPbI_3$ has high charge carrier mobility, low binding energy, high absorbance coefficient, long diffusion length for electron and hole, low trap density, and bandgap tunability. The application of MAPbI$_3$ as a perovskite layer in PSC has been reported to boost the efficiency up to 24.2% [46].

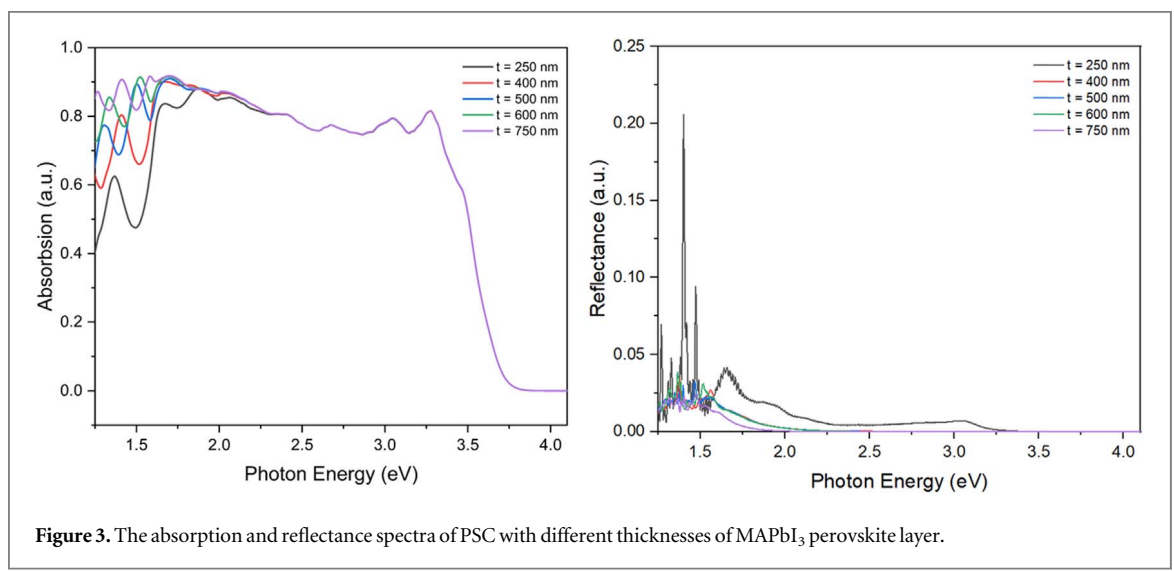

**Figure 3.** The absorption and reflectance spectra of PSC with different thicknesses of MAPbI$_3$ perovskite layer.

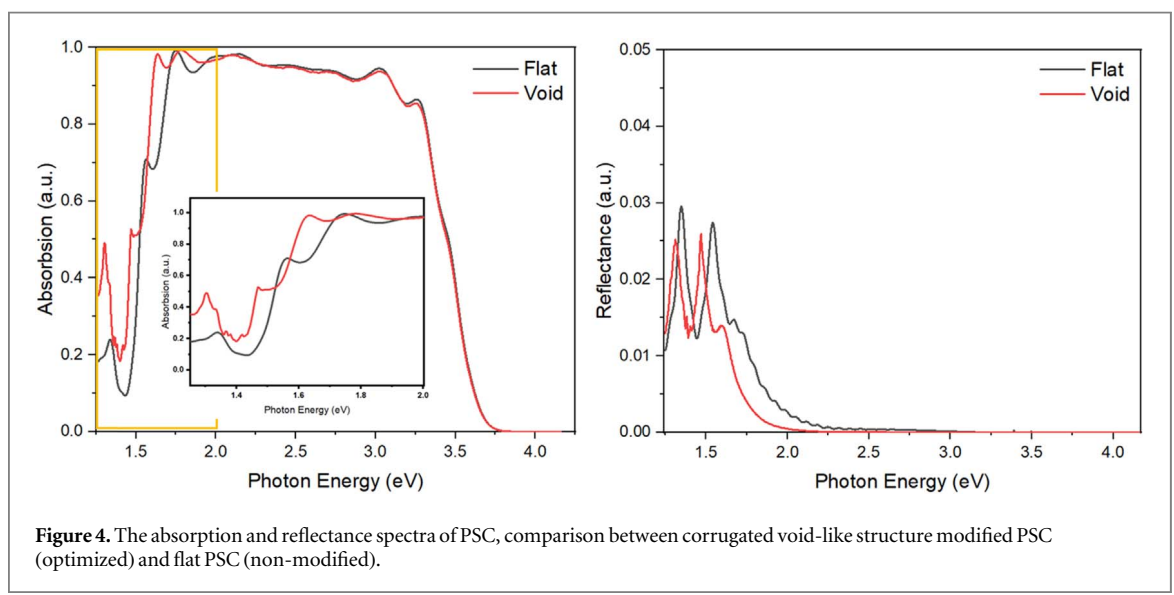

**Figure 4.** The absorption and reflectance spectra of PSC, comparison between corrugated void-like structure modified PSC (optimized) and flat PSC (non-modified).

The optimized parameters of corrugated void-like structure in PSC were applied in simulation to evaluate and compare its performance to flat design which does not apply the corrugated void-like structure. As shown in figure 4, the absorbance and reflectance performance was significantly enhanced specifically at low photon energy (1.25–2.0 eV) for void PSC compared to flat PSC. The performance enhancement is indicated by higher absorption and lower reflectance of light irradiated upon the PSC. The low photon energy region is equal to the near-infrared (NIR) region, which is the majority of the power harvested from the Sun's light. Therefore, there are several reasons for a significant increase in absorption in the NIR region. The first is due to the dominant NIR intensity in simulated Sun's rays. Second, the influence of the application of a corrugated void-like structure produces strong far-field scattering effects. The void coupling the waveguided modes trapped in the cell layers by diverting the vertically incident light, resulting in optical path length amplification. As shown in figure 5(b), in the corrugated void-like structure, the constructive and destructive interference between light waves going along the incidence direction and scattered light traveling along the plane of the cell layers experienced repeated reflections from the walls of the void structure and resulting in hotspots in the profile [33]. Lastly, You *et al* (2015) studied the charge transport properties of perovskite in contact with ETL TiO$_2$ and HTL Spiro-OMeTAD. The photoluminescence curve shows the formation of peaks at a wavelength of around 760 nm for both samples. This strengthens the reason for the significant increase in absorption in the NIR region [47]. This result has confirmed the application of corrugated void-like structure has been successfully improved the absorption of the PSC by introducing the light trapping phenomenon of the anti-reflection layer [23, 29, 30].

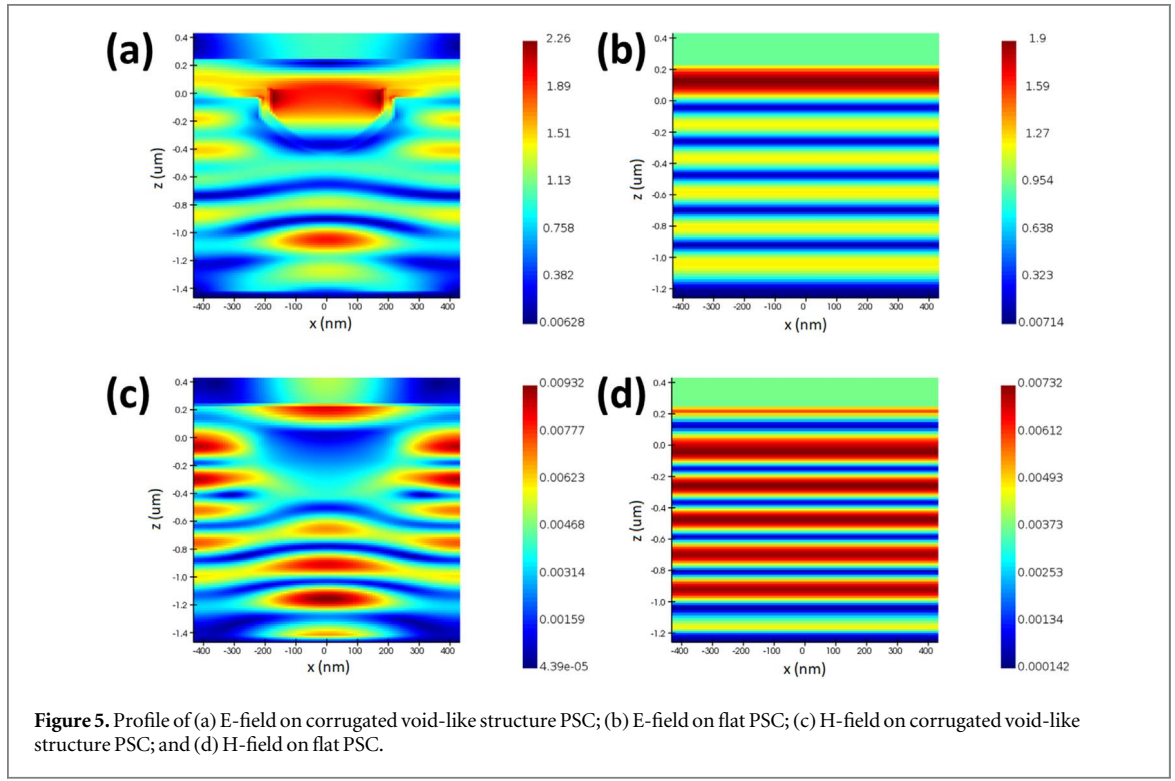

**Figure 5.** Profile of (a) E-field on corrugated void-like structure PSC; (b) E-field on flat PSC; (c) H-field on corrugated void-like structure PSC; and (d) H-field on flat PSC.

**Table 1.** Comparison of optical absorption on PSC-modified nanostructure with typical wafer structure.

| PSC Structure | Applied nanostructure | Absorption percentage | References |
| --- | --- | --- | --- |
| ITO/TiO$_2$/MaPbI$_3$/Spiro-OMeTAD/Au | Holes | 57.81% | [49] |
| ITO/TiO$_2$/MaPbI$_3$/Au | Holes | 59.28% | [48] |
| ITO/TiO$_2$/MaPbI$_3$/Spiro-OMeTAD/Ag | Void-like structure | 62.26% | [50] |
| ITO/TiO$_2$/MaPbI$_3$/Spiro-OMeTAD/Ag | Optimized void-like structure | 67.35% | This work |

The void structure effect also observed in the E-field and H-field profile of PSC structure in *z*-normal plane. As shown in figure 5, the flat structure shown high E-field only on its top layer, indicated by orange to red color. Meanwhile, the modified PSC successfully trapped the high E-field in its void structure thus resulting high E-field around the MAPbI$_3$ perovskite layer below. This E-field profile also confirmed by the H-field profile which opposite to the E-field. Refer to equation (2), the E-field $|\vec{E}|^2$ is proportional to the absorption power, $P_{Abs}$. With corrugated void-like structure, the high E-field more congregate on the front contact surface until it enters the active layer as the wavelength increases.

Compared to the reported study that applied nanostructures to PSCs, the proposed PSC with corrugated void-like structure provided superior optical performance as shown in table 1. Generally, with the typical PSC structure, void-like structures promise better optical absorption compared to conventional nanostructures in the form of holes [48–50]. This is proven by the results of this study and Haque *et al* (2019) whose absorption value reaches >60% which is higher than the absorption of hole nanostructures [50]. Furthermore, the void-like structure generated from the optimization of radius and lattice constant in this study successfully exceeded the absorption of the void-like structure proposed by Haque *et al* (2019) by ~5%. This comparison was carried out in the wavelength range of 300–1000 nm. As discussed previously, the proposed PSC is superior because it has high absorption characteristics in the low photon energy region which is equal to the NIR region. This characteristic is not owned by common PSCs whose absorption decreases significantly in the NIR region.

The optimum design of the corrugated void-like structure proposed in this study can be realized by the fabrication methods reported in several solar cell studies, for example, metal-assisted chemical etching process [51], lithography [52], and multistep anodization templating [53]. Jiang *et al* (2017) multi-crystalline silicon solar cells with inverted pyramid nanostructures. The metal-assisted chemical etching process successfully fabricates nanostructures with diameters from 50 nm to 100 nm with a 400–500 nm depth [51]. Multistep anodization templating is also used to fabricate nanostructures as nanocones for anti-reflection on PSC. The method consists of making the i-cone template by the multistep anodization, followed by PDMS drop casting on the template to form nanocone PDMS [53]. In addition, various nanostructured lithography methods have also been reported to modify crystalline silicon solar cells successfully [52]. As proposed in this study, these methods are reliable for making corrugated void-like structures in ITO materials. Furthermore, the structure fabrication on TiO2 ETL can be carried out using high-temperature annealing, spin-coating, sputter annealing, and sputter deposition methods [54].

## 4. Conclusion

In this study, theoretical work has successfully investigated the structural modification effect of the PSC light absorption phenomenon using FDTD simulation. The anti-reflection layer was modified to have a corrugated void-like structure and the thickness of the MAPbI$_3$ perovskite layer was also optimized. The modification results in high and stable absorption spectra with slight decrement throughout the visible length region and significant enhancement in NIR region. The geometry parameters of corrugated void-like structure resulting optimum absorption response at $r = 692$ nm and $a = 776$ nm. Meanwhile, the 750 nm was found to be the most suitable for the MAPbI$_3$ layer thickness. The scattering effects from corrugated void-like structure produces strong far-field, thus the electric field intensity in E-field profile of the PSC cell was found to be increasing throughout the PSC cells after introducing the structural modification. This improvement of the electric field profile indicates the enhancement of light absorption of the PSC device, which is beneficial for further application. Corrugated void-like structure-modified PSC successfully enhanced the optical absorption and has the potential for further study in electrical behavior and PCE.

## Acknowledgments

This work is supported by Universitas Pendidikan Indonesia through World Class University (WCU) Program and Directorate of Higher Education, Ministry of Education, and Culture, Research, and Technology, Republic of Indonesia with Grant Number 278/UN40.LP/PT.01.03/2021.

## Data availability statement

The data cannot be made publicly available upon publication because they are not available in a format that is sufficiently accessible or reusable by other researchers. The data that support the findings of this study are available upon reasonable request from the authors.

## ORCID iDs

Budi Mulyanti  https://orcid.org/0000-0002-4593-6289
Roer Eka Pawinanto  https://orcid.org/0000-0002-8183-1249